\documentclass[11pt,twoside]{article}
\usepackage{asp2010}

\resetcounters

\begin{document}

\title{Stellar and substellar mass function of the young open cluster
  candidates Alessi~5 and $\beta$~Monocerotis} \author{S.
  Boudreault$^{1,2,3}$ and J. A. Caballero$^{4}$ \affil{$^1$Mullard
    Space Science Laboratory, University College London, Holmbury St
    Mary, Dorking, Surrey, RH5 6NT, United Kingdom}
  \affil{$^2$Visiting Astronomer at the Department of Physics and
    Astronomy, State University of New York, Stony Brook, NY
    11794-3800, USA} \affil{$^3$ Max-Planck-Institut f\"ur Astronomie,
    K\"onigstuhl 17, D-69117 Heidelberg, Germany} \affil{$^4$ Centro
    de Astrobiolog\'{\i}a (CSIC-INTA), Departamento de
    Astrof\'{\i}sica, PO~Box~78, E-28691 Villanueva de la Ca\~nada,
    Madrid, Spain}}

\begin{abstract}
  Although the stellar and substellar populations have been studied in
  various young and old open clusters, additional studies in clusters
  in the age range from 5 to 100\,Myr is crucial (e.g. to give more
  constrains on initial mass function variation with improved
  statistics). Among the open cluster candidates from recent studies,
  two clusters are best suited for photometric survey of very-low mass
  stars and brown dwarfs, considering their youth and relative
  proximity: Alessi~5 ($\tau$$\sim$40\,Myr, $d$$\sim$400\,pc) and
  $\beta$~Monocerotis ($\tau$$\sim$9.1\,Myr, $d$$\sim$400\,pc). For
  both clusters, we performed an optical and near-infrared photometric
  survey, and a virtual observatory survey. Our survey is predicted to
  be sensitive from the massive B main sequence stars down to brown
  dwarfs of 30\,M$_{\rm Jup}$.  Here, we present and discuss
  preliminary results, including the mass function obtained for Alessi
  5, which is surprisingly very similar to the mass function of the
  Hyades ($\tau$$\sim$600\,Myr), although they are of very different
  ages.
\end{abstract}

\section{Introduction}

Several studies over the past ten years have presented surveys of open
clusters in order to study the mass function (MF) of stellar and
substellar populations, including the Orion Nebula Cluster,
$\sigma$~Orionis, IC~2391, the Pleiades and the Praesepe, to list just
a few examples. These studies are important since stars and brown
dwarfs (BD)s in open clusters possess modest age and metalicity
spreads and share a common distance, in comparisons with large
uncertainties for the field stellar and substellar objects
\citep{bastian2010}. In addition, determination of the MF in clusters
with different properties (e.g.\ different density and ages) has led
some investigators to draw conclusions about the relative efficiency
of possible BD formation mechanisms
\citep{briceno2002,chabrier2003,kroupa2003,kumar2007,boudreault2009}.

Many earlier studies of the substellar MF have focused on young open
clusters with ages less than $\sim$5\,Myr, and in many cases much
younger ($\lesssim$1\,Myr). This is partly because BDs are bright when
they are young, thus easing detection of the least massive objects.
However, youth present difficulties: intra-cluster extinction plagues
the determination of the intrinsic luminosity function from the
measured photometry, and at these ages the BD models have large(r)
uncertainties \citep{baraffe2002}. Studies in older clusters
($\gtrsim$100\,Myr) present difficulties too: lacking a significant
nuclear energy source, BDs cool and get faint as they age, so deeper
surveys are required to detect them, and low-mass objects evaporate
from clusters by dynamical evaporation
\citep{marcos2000,adams2002b,bouvier2008}.  Clusters in the range of
age 5--100\,Ma are perfect tools for MF studies on the BDs and very
low-mass star populations since (1) these objects are bright,
(2) have not lost trace of initial condition due to dynamical mass
segregation or dynamical evaporation and (3) low extinction is
expected towards these clusters. Despite these advantages, only a few
open clusters are known in this age range.

Some works have been performed to search for previously unknown open
clusters. Among the open cluster candidates from \citet{alessi2003}
and \citet{kharchenko2005}, two clusters are best suited for
photometric survey of very low-mass star and BDs considering their
youth and relative proximity\,: Alessi~5 ($\tau$\,$\sim$40\,Myr,
$d$\,$\sim$400\,pc; \citealt{alessi2003}) and $\beta$~Monocerotis
($\tau$\, $\sim$ 9\,Ma, $d$\, $\sim$ 400\,pc;
\citealt{kharchenko2005}).  (This cluster is presented as ``ASCC~24''
in \citet{kharchenko2005}.) So far, no accurate studies of these two
clusters have been performed.

\section{\label{observations} Observations}

\subsection{\label{optical} Optical photometry: WFI observations, data
  reduction, astrometry and photometric calibration}

Our survey consists of one single Wide Field Imager (WFI) field of
size 34$\times$33 arcmin$^2$, observed in wide band $R_{\rm c}$, and
medium band 770/19, 815/20, 856/14 and 914/27 (where the filter name
notation is central wavelength on the full width at half maximum,
FWHM, in nm). This gives a total coverage of 0.26\,deg$^2$ observed in
all five bands, centred on the brightest stars of each cluster
candidate.  The data reduction and photometric calibration was
performed in a similar way as presented in \citet{boudreault2009}. To
correct for Earth-atmospheric absorption on the photometry, we
observed the spectrophotometric standard stars observed were LTT~3864
and 4364.

For our $\beta$~Mon and Alessi~5 observations, the 5$\sigma$ detection
limits of our survey are $R_{\rm c}$=22.9\,mag, which correspond to
$\sim$30\,M$_{\rm Jup}$ according to our dust-free isochrone. The root
mean square accuracy of our astrometric solution was 0.15--0.20.

\subsection{\label{nir} Near infrared photometry: $\Omega$2k and
  and 2MASS}

The near infrared observations were performed only for $\beta$~Mon.
There were made using four Omega 2000 ($\Omega$2k) pointing on the
3.5m telescope at Calar Alto, Spain, with observation runs of several
nights in December 2008 and December 2009, covering the WFI pointing.
The data reduction and photometric calibration of our near infrared
photometry with $\Omega$2k was performed in a similar way as presented
in \citet{boudreault2010}. The 5$\sigma$ detection limit at
$J$=20.7\,mag corresponds to an object of $\sim$30\,M$_{\rm Jup}$ in
$\beta$~Mon.  For our survey on Alessi~5, we used the $J$ and $K_{\rm
  s}$ photometry from 2MASS.

\section{\label{selection} Candidate Selection Procedure}

The procedure to compute the masses and effective temperature based on
photometry is done in a similar way as presented by
\citet{boudreault2009} and \citet{boudreault2010}. 

In order to perform the selection of candidates, we compute an
isochrones for both Alessi~5 and $\beta$~Mon. We used the spectral
energy distribution to derive the mass and effective temperature,
$T_{\rm eff}$, assuming that all our photometric candidates belong to
the clusters studied. We used evolutionary tracks from
\citet{baraffe1998} and atmosphere models from \citet{hauschildt1999a}
(assuming a dust-free atmosphere; the NextGen model) to compute an
isochrone for Alessi~5 using an age of 40\,Myr, distance of 400\,pc, a
solar metalicity and neglecting the reddening ($E(B-V)$\,=\,0\,mag),
and for $\beta$~Mon using an age of 9.1\,Myr, distance of 210\,pc, a
solar metalicity and neglecting the reddening.

\begin{figure}[!ht]
\plotfiddle{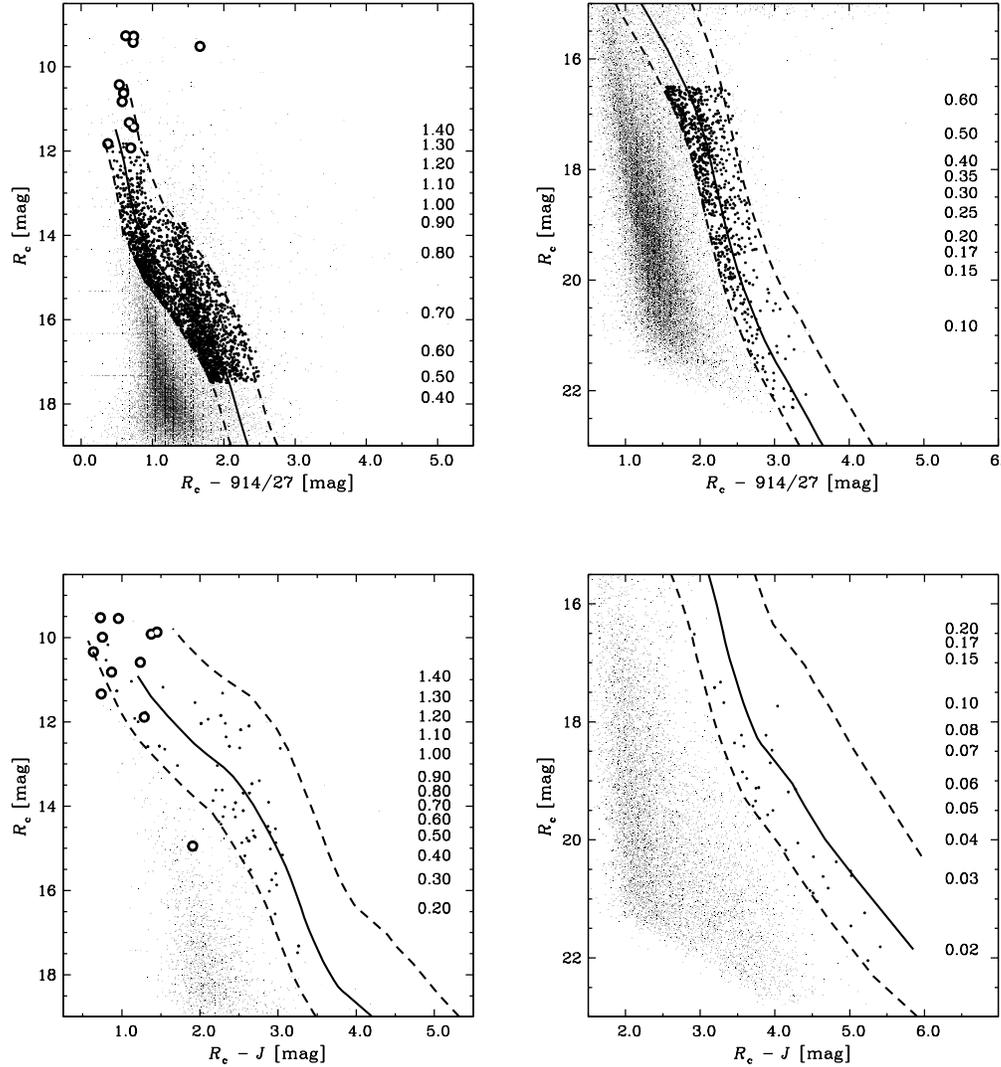}{14.0cm}{0}{70}{70}{-225}{-110}
\caption{\label{fig:cmd-nextgen} Colour--magnitude diagram for
  Alessi~5 (~\textit{top two panels}~) with the $R_{\rm c}$ and 913/27
  bands used in the selection procedure, and for $\beta$~Mon
  (~\textit{top two panels}~) with the $R_{\rm c}$ and $J$ bands. We
  present the colour--magnitude diagrams from our shallow images
  (~\textit{left panels}~) and deep images (~\textit{right panels}~)
  we have taken. As solid lines we show the isochrone computed from an
  evolutionary model with a dust-free atmosphere (NextGen model).  The
  numbers indicate the masses (in M$_{\odot}$) on the model sequence
  for various $R_{\rm c}$ magnitudes.  We also show candidate cluster
  members that we detected in our survey from \citet{kharchenko2005}
  (~\textit{circles}~). The dashed lines delimit our selection band.}
\end{figure}

Candidates were first selected from the CMD involving the wide band
$R_{\rm c}$ and $913/27$ for Alessi~5, and the wide bands $R_{\rm c}$
and $J$. The candidates are only objects within a selection area
defines to include (1) error on the distance, (2) error on the age,
(3) error on the photometry and (4) objects brighter than the
isochrones by 0.753\,mag in order to include unresolved binaries.  In
Fig.~\ref{fig:cmd-nextgen} we present the CMDs where candidates were
selected. The Fig.~\ref{fig:cmd-nextgen} also show cluster member of
\citet{kharchenko2005} with a membership probability higer than 10\%,
based on proper motion.

\begin{figure}[!ht]
\plotone{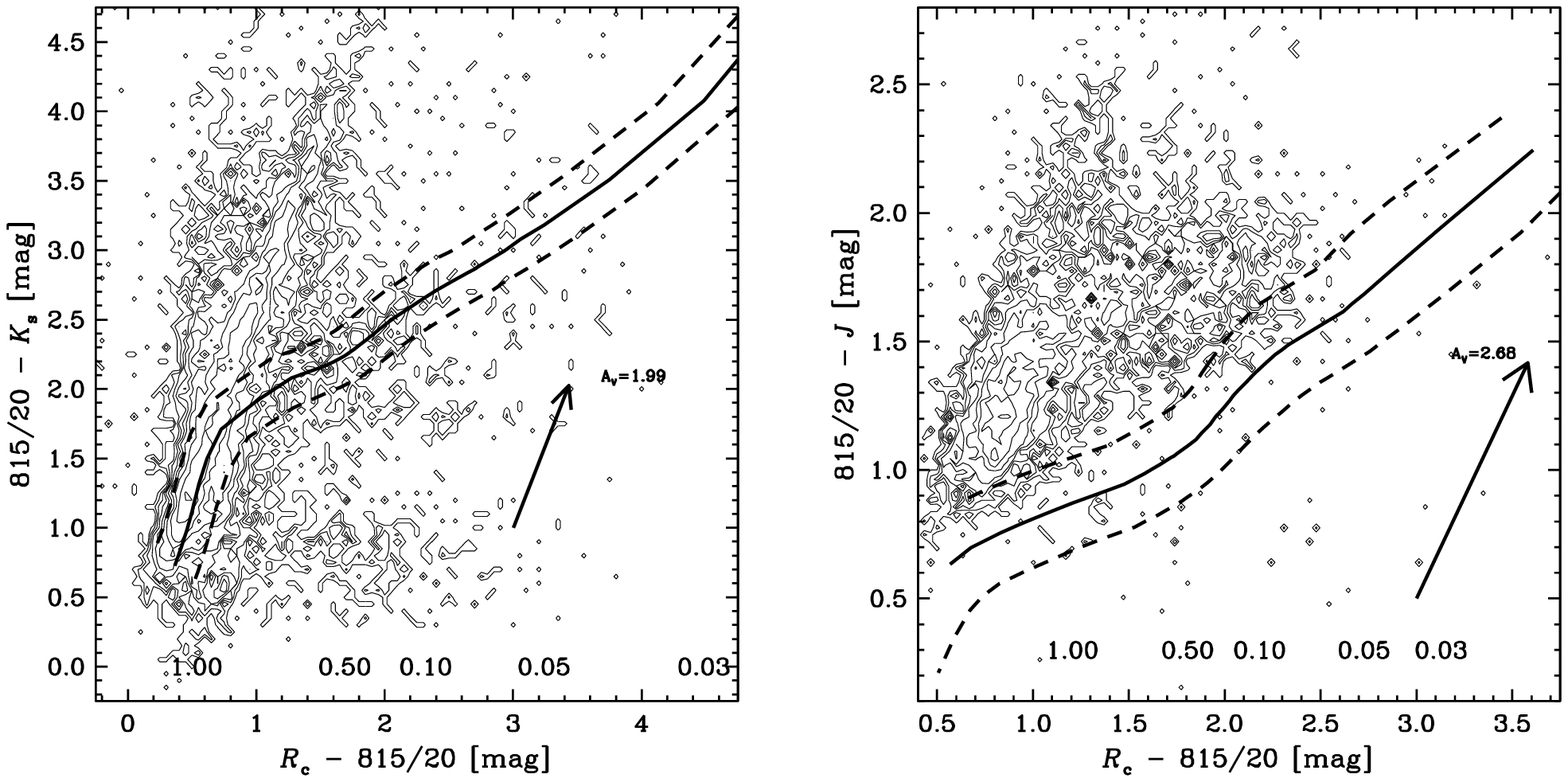}
\caption{\label{fig:ccd-contour} Colour--colour diagram for Alessi~5
  (~\textit{left panel}~) with the $R_{\rm c}$, 815/20 and $K_{\rm s}$
  bands used in the selection procedure, and for $\beta$~Mon
  (~\textit{right panel}~) with the $R_{\rm c}$, 815/20 and $J$ bands.
  As for Fig.~\ref{fig:cmd-nextgen}, as solid lines we show the
  isochrone computed from an evolutionary model with a dust-free
  atmosphere and the dashed lines delimit our selection band. For
  Alessi~5, we clearly see a structure overlaping the isochrones for
  masses lover than $\sim$0.6\,M$_\odot$. On the other hand, we don't
  observe any structure in the colour--colour diagram of $\beta$~Mon
  that overlap with the isochrone.}
\end{figure}

The second stage of candidate selection was achieved using
colour-colour diagrams using the $R_{\rm c}$, 815/20 and $K_{\rm s}$
bands for Alessi~5, and using the $R_{\rm c}$, 815/20 and $J$ bands
for $\beta$~Mon (Fig.~\ref{fig:ccd-contour}). Since colours are used
here, the selection area is defined by the error on the age on the
isochrones and by the error on the photometry. For clarity, we present
a contour plot of the colour--colour diagram for both clusters.

As indicated previously, our determination of $T_{\rm eff}$ is based
on the spectral energy distribution of each object and is independent
of the assumed distance. The membership status of an object can
therefore be assessed by comparing its observed magnitude in a band
with its magnitude predicted from its $T_{\rm eff}$ and $\beta$~Mon
and Alessi~5's isochrone (which assumes a distance and an age). This
selection step is only a verification of the consistency between the
physical parameters obtained of the photometric cluster candidates
with the physical properties assumed for the cluster itself when
computing the isochrones. To avoid removing unresolved
binaries that are real members of the cluster, we keep all objects
with a computed magnitude of up to 0.753\,mag brighter than the
observed magnitude.

\section{\label{results} Results and discussions}

\subsection{\label{results-alessi5} Alessi~5}

\begin{figure}[!ht]
\plotfiddle{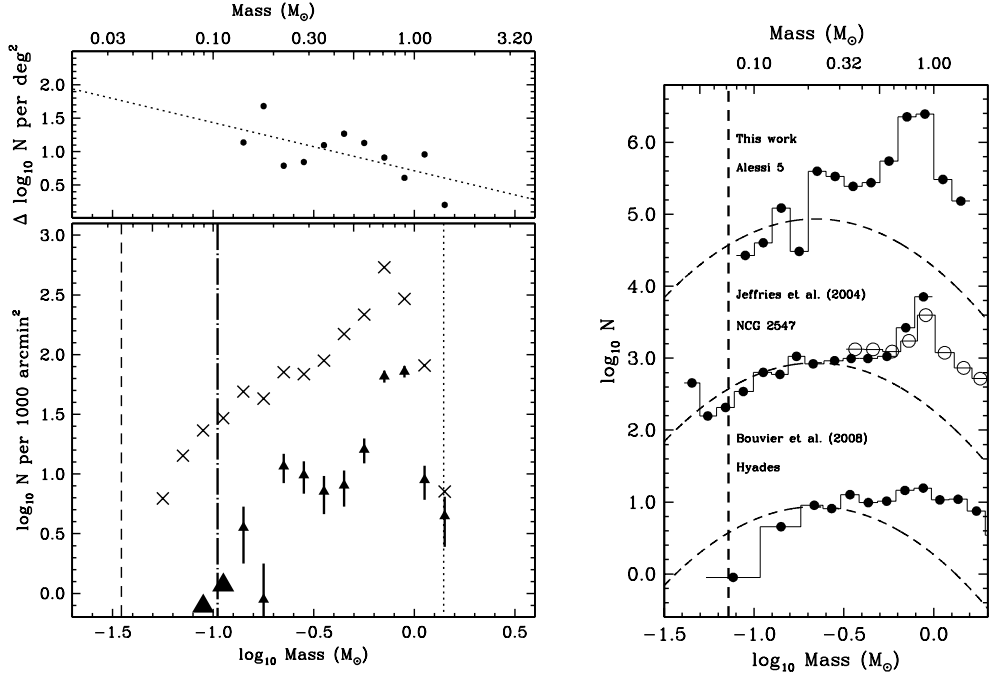}{8.5cm}{0}{75}{75}{-210}{-130}
\caption{\label{fig:mf} \textit{Lower left panel.} MF of Alessi~5
  using only our optical photometry (\textit{crosses}) and combined
  with the near infrared photometry of 2MASS (\textit{triangles}). The
  vertical dotted line represents the saturation limit of our optical
  survey, the vertical dashed line represent the 5\,$sigma$ detection
  limit of our optical survey, and the vertical dash-dotted line
  represents the 10\,$\sigma$ detection limit of 2MASS. Error bars are
  Poissonian arising from the number of objects observed in each bin.
  \textit{Top left panel.}  Difference of the number of object, in
  each mass bin, between the MF computed using our optical photometry
  and the MF computed using the combination of the optical data from
  WFI and the near infrared $JK_{\rm s}$ data from 2MASS.
  \textit{Lower left panel.} MF of Alessi~5 and of the Hyades and
  NGC~2547.  We also show the galactic field star MF fit from
  \citet{chabrier2003} as a thin dashed line and the substellar limit
  as a thick dashed line. We have normalized all the MFs to the
  log-normal fit of \citet{chabrier2003} at $\sim$0.3\,M$_\odot$
  (log\,M=-0.5).}
\end{figure}

We obtain a total of 234 cluster candidates based on our deep
photometric survey. The $J$ and $K_{\rm s}$ photometry of 2MASS is
shallower in terms of mass in Alessi~5 compared to our optical
photometry. To compute the MF of Alessi~5 to the lowest mass bin
reached without optical data, we have computed a MF using only the
optical photometry with WFI. We present this MF on Fig.~\ref{fig:mf}
(lower left panel) as crosses. We computed a second MF from the list
of candidates that passes all selections criteria with near infrared
$J$ and $K_{\rm s}$ photometry from 2MASS (presented on
Fig.~\ref{fig:mf}, lower left panel, as filled triangles).  For each
mass bin, we computed the number of object removed by adding the $J$
and $K_{\rm s}$ photometry of 2MASS to our selection process and mass
determination (this is plotted as a function of mass in
Fig.~\ref{fig:mf}, top left panel). We fitted a power-law function to
estimate the number of object that would be removed \textit{if} we
would had additional $J$ and $K_{\rm s}$ photometry added to our
optical photometry. The corresponding extension of the MF is given as
large triangles (Fig.~\ref{fig:mf}, lower left panel).

In Fig.~\ref{fig:mf} (right panel) we compare the MF of Alessi~5 with
the MF of the open cluster NGC~2546 and the Hyades. The MF of Alessi~5
is surprisingly similar to the MF of the Hyades \citep{bouvier2008}
with a decrease in the MF below $\sim$0.6\,M$_\odot$, although they
are of very different ages (with about 40 and 600\,Myr respectively).
This support the conclusion that initial conditions are more likely to
influence the shape of the MF more significantly than dynamical
evolution. In addition, the MF of Alessi~5 shows some similarities
with the MF of NGC~2547 \citep{jeffries2004}: both show a decrease in
the MF below $\sim$0.6\,M$_\odot$ and both open clusters present a
peak at 0.7--1.0\,M$_\odot$. It was shown by \citet{boudreault2009}
that this peak is due to red giant background contaminants.

\subsection{\label{results-betamon} $\beta$~Mon}

We obtain a total number of object of 19 candidates from our deep
photometric survey in $\beta$~Mon. Considering this and the absence of
any structure in the colour-colour diagram of $\beta$~Mon (see
Fig.~\ref{fig:ccd-contour}), we conclude that were is no cluster
towards $\beta$~Mon. This is further confirmed with our virtual
observatory study in the following section.

\section{\label{vo} Virtual Observatory study}

\begin{figure}[!ht]
\plotfiddle{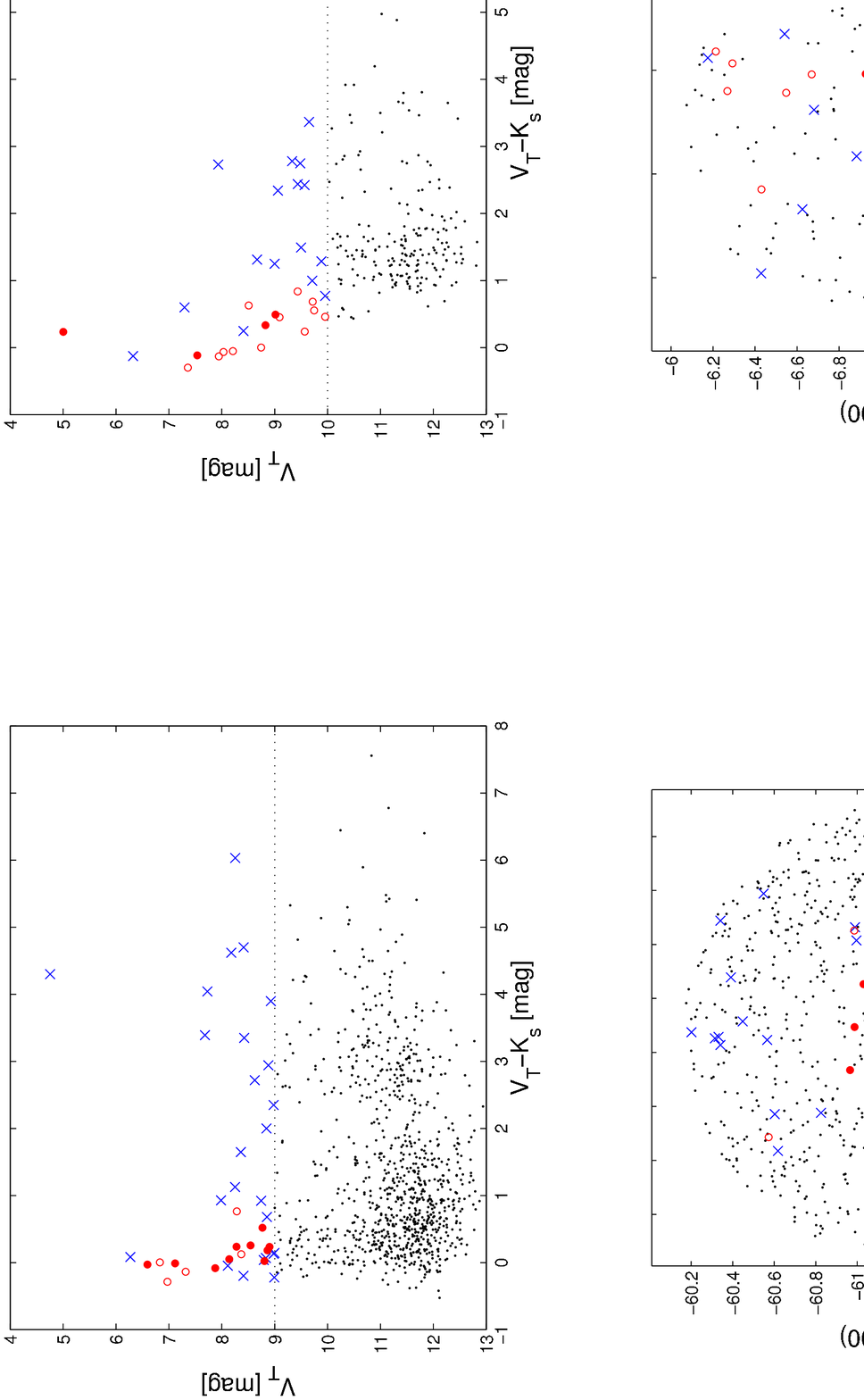}{9.9cm}{-90}{45}{45}{-245}{+350}
\caption{\label{fig:cmd-ra-dec-vo} Near infrared-optical CMD of
  Alessi~5 (\textit{top left panel}) and of $\beta$~Mon (\textit{top
    right panel}) using $V_{\rm T}$ and $K_{\rm S}$ band from our
  virtual observatory study.  Spatial location diagrams of the
  cross-matched Tycho-2 and 2MASS sources in the 1\,deg-radius
  circular areas centred on HD~93010~A (in the centre of Alessi~5,
  \textit{lower left panel}), and $\beta$~Mon~ABC (\textit{lower right
    panel}). In all the panels, (red) filled circles are cluster
  member candidates, (red) open circles are other early-type stars in
  the region, (blue) crosses are non-members based on abnormal proper
  motions and/or colours, and (black) small dots are the remaining
  cross-matched sources. For the CMDs, the horizontal dotted lines
  indicate the size of the virtual observatory analysis.}
\end{figure}

We performed a 6.3\,deg$^2$ Aladin-based virtual observatory analysis
of the high-mass (about 8 to 1\,M$_\odot$) population of stars in the
Tycho-2 catalogue over both clusters. We use near infrared-optical
CMDs (presented in Fig.~\ref{fig:cmd-ra-dec-vo}, top two panels),
proper-motions, spatial location diagrams of the cross-matched Tycho-2
and 2MASS sources in the 1\,deg-radius circular areas centred on
HD~93010~A for Alessi~5 and on $\beta$~Mon~ABC for $\beta$~Mon
(presented in Fig.~\ref{fig:cmd-ra-dec-vo}, bottom two panels), and
normalized cumulative distributions. At a quick glance to the spatial
distribution diagrams in Fig.~\ref{fig:cmd-ra-dec-vo}, the
``clustering'' of Alessi~5 is obvious, with ten cluster members in the
innermost 20\,arcmin-radius circular area and only five early-type
stars possibly not associated to the cluster. However, $\beta$~Mon has
only four cluster member candidates in the same 20\,arcmin-radius
circular area and up to twelve early-type stars homogeneously
distributed in the corona between 40 and 60\,arcmin to
$\beta$~Mon~ABC.

From our virtual observatory studies, including our above analysis of
the spatial distribution of the cluster candidates, we conclude that
there is no real clustering around $\beta$~Mon (i.e.  {\em
  [KPR2004]~24 does not exist}), but confirm the existence of Alessi~5
around the early-type giant binary HD~93010~AB.

\section{\label{conclusions} Conclusions}

In this proceeding we presented the results of a survey to identify
high- to low-mass stars and brown dwarf members of the recently
discovered open cluster candidates Alessi~5 and $\beta$~Mon. Our
survey consisted of an optical and near infrared photometric survey
covering 0.26\,deg$^2$ and a virtual observatory survey of
6.3\,deg$^2$ for both Alessi~5 and $\beta$~Mon. With a 5$\sigma$
detection limits of $R_{\rm c}$=22.9\,mag, our survey is predicted to
be sensitive from the massive B main sequence stars down to brown
dwarfs of 30\,M$_{\rm Jup}$ in Alessi~5 and in $\beta$~Mon.

From our optical observations of Alessi~5 we identify 234 low-mass
cluster member candidates from our WFI+2MASS survey and 10 high-mass
candidates from our Aladin-based survey. From the list of candidates,
we computed the MF of this cluster. The MF of Alessi~5 is surprisingly
similar to the one of the Hyades, although they are of very different
ages (with about 40 and 600\,Myr respectively).  This support the
conclusion that initial conditions are more likely to influence the
shape of the MF more significantly than dynamical evolution. In
addition, the MF of Alessi~5 shows some similarities with the one of
of NGC 2547.

As for the open cluster $\beta$~Mon, we report a non detection of any
clustering at the distance surveyed.

The results reported here will be presented with further details in an
future publication (Boudreault \& Caballero 2010, in prep).

\acknowledgements S.B. acknowledge support from the Deutsche
Forschungsgemeinschaft (DFG) grant BA2163 (Emmy-Noether Program) to
Coryn A. L.  Bailer-Jones. Partial financial support was provided by
the Spanish Ministry of Science under grant AyA2008-06423-C03-03.
Some of the observations on which this work is based were obtained
during ESO programmes 081.A-9001(A).

\bibliography{boudreault_s}

\end{document}